\documentclass[%prl,
conference]{IEEEtran}
\IEEEoverridecommandlockouts
  \usepackage{caption}
\usepackage{textcomp}
\usepackage{xcolor}
\def\BibTeX{{\rm B\kern-.05em{\sc i\kern-.025em b}\kern-.08em
    T\kern-.1667em\lower.7ex\hbox{E}\kern-.125emX}}
\usepackage{tikz}
\usepackage{bbm}
\usepackage{cite}
\usepackage{adjustbox}
\usetikzlibrary{arrows, hobby}
\usepackage{wrapfig}
\usepackage{graphicx}
\usepackage{amsmath}
\usepackage{amssymb}
\usepackage{hyperref}
\usepackage{ mathrsfs }
\usepackage{indentfirst}
\usepackage{braket}
\usepackage{color}
\usepackage{empheq}
\usepackage{lipsum,adjustbox}
\usepackage[outdir=./]{epstopdf}
\usetikzlibrary{calc,positioning}
\tikzstyle{beamsplitter}=[fill=blue, fill opacity=0.2]

\usepackage[english]{babel}
\usepackage[utf8]{inputenc}
\begin{document}

\title{Use of a Local Local Oscillator for the
Satellite-to-Earth Channel}
\author{
	\IEEEauthorblockN{S. P. Kish$^1$, E. Villase\~nor$^1$, R. Malaney$^1$, K. A. Mudge$^2$ and K. J. Grant$^2$}\\
	\IEEEauthorblockA{${}^1$School of Electrical Engineering  \& Telecommunications,\\
		The University of New South Wales, Sydney, NSW 2052, Australia.\\
		${}^2$Cyber and Electronic Warfare Division, \\
		Defence Science and Technology Group, Edinburgh, SA 5111, Australia.}\\
}
%	
%\author{\IEEEauthorblockN{1\textsuperscript{st} Sebastian Kish}
%\IEEEauthorblockA{\textit{School of EE\&T} \\
%\textit{University of New South Wales}\\
%Sydney, Australia \\
%s.kish@unsw.edu.au}
%\and
%\IEEEauthorblockN{2\textsuperscript{nd} Eduardo Villase\~nor}
%\IEEEauthorblockA{\textit{School of EE\&T} \\
%\textit{University of New South Wales}\\
%Sydney, Australia  \\
%e.villasenoralvarez@student.unsw.edu.au}
%\and
%\IEEEauthorblockN{3\textsuperscript{rd} Robert Malaney}
%\IEEEauthorblockA{\textit{School of EE\&T} \\
%\textit{University of New South Wales}\\
%Sydney, Australia  \\
%r.malaney@unsw.edu.au}
%\centering
%\linebreak
%\and
%\IEEEauthorblockN{\centering 4\textsuperscript{th} Kerry A. Mudge}
%\IEEEauthorblockA{\textit{Cyber and Electronic Warfare Division} \\
%\textit{Defence Science and Technology Group}\\
%Edinburgh, Australia  \\
%kerry.mudge@dsto.defence.gov.au}
%\and
%\IEEEauthorblockN{5\textsuperscript{th} Kenneth J. Grant}
%\IEEEauthorblockA{\textit{Cyber and Electronic Warfare Division} \\
%\textit{Defence Science and Technology Group}\\
%Edinburgh, Australia  \\
%ken.grant@dst.defence.gov.au}}
%%\author{ S. P. Kish}\affiliation{School of Electrical Engineering and Telecommunication, \\
%%	University of New South Wales, Sydney, New South Wales 2052, Australia}

\date{\today}
\maketitle
\begin{abstract}
Continuous variable quantum key distribution ($\text{CV-QKD}$) offers information-theoretic secure key sharing between two parties. The sharing of a phase reference frame is an essential requirement for coherent detection in CV-QKD. Due to the potential attacks related to transmitting the local oscillator (LO) alongside quantum signals, there has been a focus on using local LOs (LLOs) to establish a shared phase reference. In this work, we develop a new noise model of a current state-of-the-art LLO scheme in the context of the satellite-to-Earth channel. In doing this, we encapsulate detailed phase-screen calculations that determine the coherent efficiency - a critical parameter in free-space CV-QKD that characterizes the wavefront aberrations caused by atmospheric turbulence. Using our new noise model we then determine the CV-QKD key rates for the satellite-to-Earth channel, secure under general attacks in the finite-size regime of the LLO scheme. Our results are of practical importance for next-generation quantum-enabled satellites that utilize multi-photon technology as opposed to single-photon technology.

\end{abstract}

%
%\pacs{03.67.Hk, 06.20.-f, 84.40. Ua}

%\maketitle

%\tableofcontents

\vspace{10 mm}
\

%\section{Discrete variable (DV) protocols}
%
%How can we characterize loss in multimode single photon wavepackets in the BB84 protocol? When can we treat them as single modes?
\section{Introduction}
%\subsection*{Local local oscillators}

Quantum key distribution (QKD) offers information-theoretic secure key distribution between two parties \cite{bed}. However, it is uncertain which of the two versions of QKD- discrete-variable (DV) QKD using single-photon technology or continuous-variable (CV) QKD using multi-photon technology will prevail.
CV-QKD is appealing because it can be implemented with current off-the-shelf technology \cite{laud}. However, accurate and precise phase recovery is required for the coherent detection of CV-QKD protocols \cite{gunt}. This is particularly the case since the sharing of a phase reference is an essential requirement in CV-QKD, in order to sift signals to information bits. The subsequent phase noise associated with the phase recovery contributes to the excess noise $\xi$- an important parameter determining the performance of $\text{CV-QKD}$. % However, it is known that the transmitted LO (TLO) design is susceptible to attacks, limiting the feasible distance for secure key distribution \cite{qi, soh2015, huang}.

The traditional and simplest implementation (which we refer to as the transmitted local oscillator (TLO) scheme) of establishing a shared phase reference is the transmission of the local oscillator (LO) from Alice to Bob which acts as a fixed phase reference for the quantum signal detection. However, the TLO scheme is not without issues, as an eavesdropper can in principle obtain access to the LO, modify it, and subsequently obtain information on the quantum key. Attacks of this form on the LO have been extensively studied including equal-amplitude attacks \cite{ma}, wavelength attacks \cite{hchrist} and calibration attacks \cite{paul, jzhuang}. Another disadvantage of the TLO scheme is that the LO is attenuated during channel tranmission, and shot-noise limited coherent detection may not be attained for lossy channels \cite{huang}.

Recently, there has been a focus on using \textit{local local oscillators} (LLO) for which the security issues of sending the LO are eliminated\footnote{We note, that even though the security aspects of an LLO appear intuitively attractive- no formal security analysis on par with known CV-QKD information-theoretic proofs (which do not consider phase referencing issues) is available. Consideration of the formal security for LLO-based protocols and system models under circumstances where the eavesdropper prepares ancillary states that become entangled with both  reference pulses and quantum signals would be useful in this regard.} by generating the LO locally at Bob's trusted device \cite{huang}. Unlike the TLO scheme, LLO schemes do not require phase or frequency lock ahead of time. In one use of an LLO, a scheme is proposed where reference pulses (or pilot tones) are sent with the signal \cite{huang}. We refer to these sequential schemes as the S-LLO scheme which was proposed in \cite{huang} and demonstrated experimentally in 25~km optical fibre independently in \cite{qi} \& \cite{soh2015}. In this scheme, two lasers are used, one at Alice for generating the quantum signal and another at Bob for the LLO. To establish a common phase reference, Alice sends low intensity reference pulses (RPs) to estimate the phase and correct the signal \cite{huang}. However, additional excess noise is introduced by the phase estimation process. Since the LLO and signal are not phase locked, there is considerable {\it phase drift} caused by the de-synchronized lasers in addition to the quantum-limited phase noise.% In the S-LLO scheme, there is also photon leakage from the reference pulse to the signal due to the amplitude modulator, contributing to the excess noise \cite{qi, soh2015}.  

%In such systems, two lasers are used, one at Alice for generating the quantum signal and another at Bob for the LLO. Reference pulses are sent with the signals which are used to calibrate the phase offset in the LLO \cite{ren, koashi, huang, marie}. However, the phase offset derived from the reference pulses is limited by quantum uncertainty, and the two lasers necessarily introduce some phase drift. In comparison to optical fibre, no phase compensation method exists for the LLO in FSO channels where additional phase noise due to atmospheric turbulence is expected \cite{qi}, but these are not insurmountable issues. Despite their intuitive appeal over LO schemes, attack strategies against LLO schemes do exist. In one such attack recently proposed \cite{wzhao}, Eve intercepts the reference pulse and estimates the phase drift using Bayesian algorithms. Consequently, the excess noise is biased to mask an intercept-resend attack and a beam-splitter attack.

The phase drift noise contribution to the excess noise is one of the drawbacks in practical implementation of the S-LLO scheme \cite{qi, soh2015, huang}.  Regardless of these practical issues, there are fundamental security issues associated with this phase drift noise, which opens up the S-LLO scheme to attacks by an eavesdropper \cite{wzhao, huang19, ren2}. Recently, much effort has been directed to minimizing phase drift using phase compensation methods in the S-LLO scheme \cite{yguo, mzou}. However, a design proposal by \cite{marie} called the delay-line LLO (D-LLO), uses balanced interferometers to eliminate the phase drift by ensuring self-coherence between signal and reference pulse. Further improvement of the D-LLO was demonstrated in optical fibre \cite{multiplex} \& \cite{wang18} by using multiplexing techniques to reduce photon leakage from the reference to signal pulse. In this work, we consider the D-LLO scheme to be a current state-of-the-art LLO scheme. However, it is not known how the D-LLO scheme would be adapted to the satellite-to-Earth channel. %In this work, we will investigate the key rates of a practical D-LLO scheme in the satellite-to-Earth channel.%unconditional security for the coherent state QKD protocol has yet to be proven. 

%Additionally, improvement of this scheme using non-Gaussian quantum photon catalysis operations has been proposed \cite{ye}. 
%Additionally, recent work had also proposed phase reference sharing without the use of a reference pulse at all (\cite{ren2, su2019}) where publicly announced raw key values are invoked to determine the phase offsets to be applied to the LLO. However, currently these types of phase determinations for the LLO are not formally proven secure, and rely on classical encryption techniques. $ye, ren2, su2019

The contributions of this work are the following: 
\begin{itemize}
\item We develop a practical noise model of a current state-of-the-art LLO scheme (the D-LLO scheme) in the satellite-to-Earth channel.
\item We numerically simulate the wavefront aberrations characterized by the coherent efficiency $\gamma$ in the satellite-to-Earth channel with and without adaptive optics (AO). 
\item We calculate for the first time the expected information-theoretic secure key rates under general attacks in the finite-size regime of the D-LLO scheme in the satellite-to-Earth channel. 
\end{itemize}

%We find that the secret key rate under general attacks in the finite-size regime is impacted by wavefront aberrations, necessitating additional AO components to obtain improved key rates. 
%In particular, the D-LLO under minimal security assumptions outperforms the S-LLO and TLO schemes. 

The remainder of this article is organized as follows. In Section \ref{noisemodel}, we adapt the D-LLO scheme from \cite{multiplex} to the satellite-to-Earth channel. In this same section, we introduce the noise model, including contributions due to the turbulent atmosphere. 
In the same section, we introduce analytic solutions for the reference pulse intensity to minimize the excess noise in the D-LLO scheme. In Section \ref{experiment}, we simulate $\gamma$ in the satellite-to-Earth channel with and without AO using techniques found in \cite{eduardo}. In Section \ref{secretkeyrate}, we calculate the achievable key rates under general attacks in the finite-size regime of the D-LLO scheme in the satellite-to-Earth channel with the values of $\gamma$. %In particular, we discuss in Section \ref{unconditional} other attacks that cannot guarantee unconditional security of the S-LLO and TLO schemes. 
Lastly, we summarize the article and discuss future directions in the conclusion.%Thirdly, we discuss potential 

\section{Noise model for LLO CV-QKD in the Satellite-to-Earth channel}
\label{noisemodel}
In the {\it free-space} optical (FSO) channel CV-QKD, contributions to the excess noise are quite different from those seen in optical fibre \cite{qi, soh2015, ren, huang, marie}. In particular, there are studies of TLO CV-QKD protocols in FSO channels that suggest excess noise due to time-of-arrival fluctuations caused by atmospheric turbulence cannot be ignored \cite{twang, deq, mli, kish1}. Fluctuations of the pulse intensity due to scintillation also contributes to the excess noise $\xi$. In \cite{marie}, these terms due to the atmospheric channel in LLO schemes were not taken into account. 

Unlike the TLO scheme, {\it wavefront aberrations} caused by propagation through atmospheric turbulence of the signal contributes to the excess noise in the LLO schemes \cite{eduardo}. This excess noise contribution is characterized by the coherent efficiency $\gamma$, determined by interfering the signal and the LO at the coherent detector. It is well known in coherent classical communication, that AOs can correct wavefront aberrations and significantly decrease the bit error rate \cite{hjian}. Recently, performance improvements using AO have been shown to improve key rates under collective attacks in the asymptotic limit in CV-QKD systems \cite{ywang, geng}. However, the impact of $\gamma$ on the secret key rate under general attacks in the deployable setting of the finite regime is yet to be investigated.
\subsection{System model}
\begin{figure*}[t!]
\centering
\includegraphics[scale=0.44]{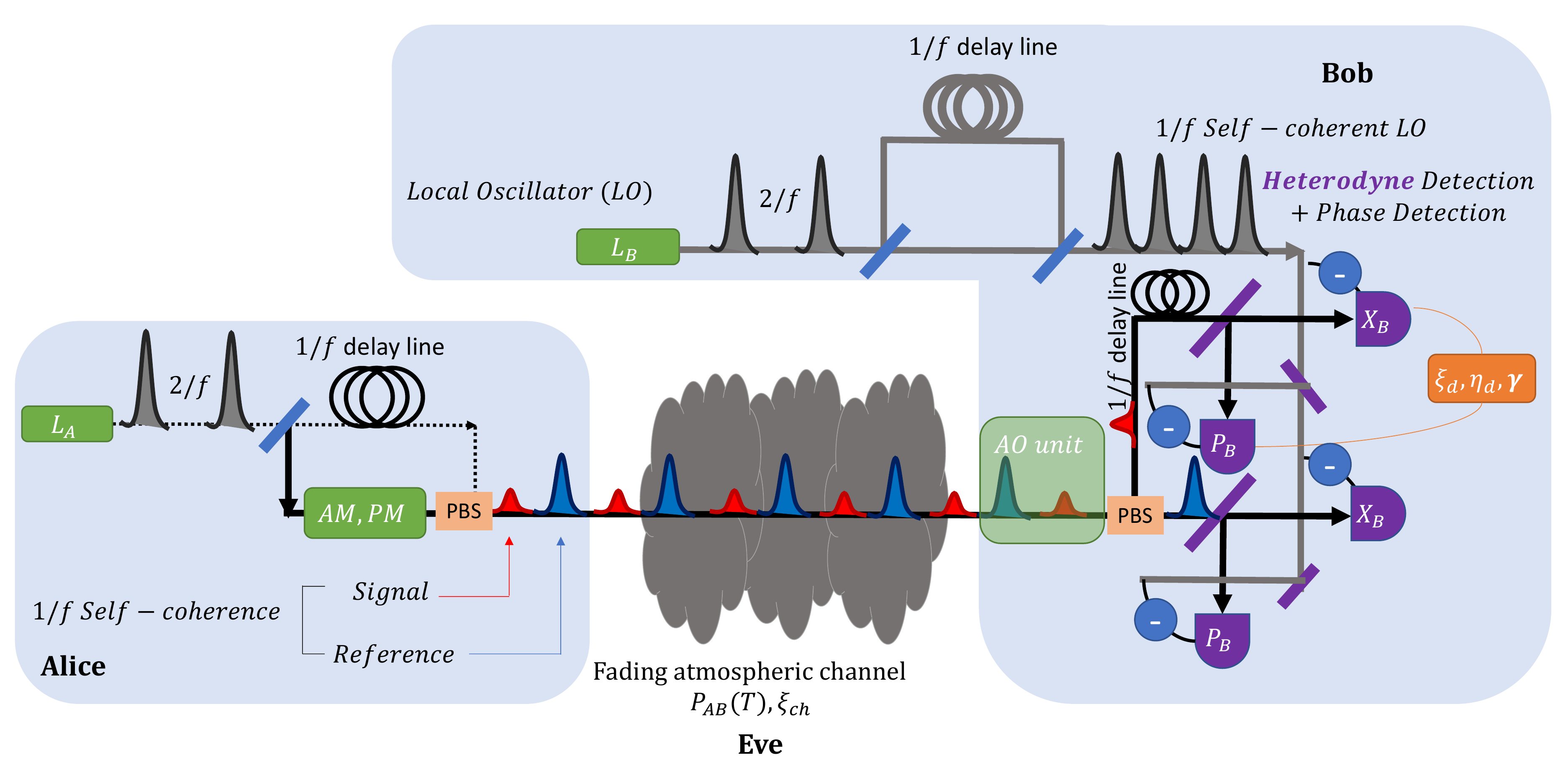}
\caption{The D-LLO protocol in the satellite-to-Earth channel. Alice's laser $L_A$ generates pulses separated by $2/f$ where $f$ is the repetition rate. A balanced interferometer is used to create self-coherent signal and reference pulses delayed by $1/f$. The signal passes through the amplitude modulator (AM) and phase modulator (PM). The reference and signal pulses are recombined with a polarizing beam-splitter (PBS) and pass through a lossy channel of transmissivity $T$ and channel excess noise $\xi_{ch}$. For the LO, pulses generated by laser $L_B$ pass through a balanced interferometer to create self-coherent pulses delayed by $1/f$. Bob receives the signal and the reference pulse which are separated. One of the two heterodyne detectors is used to determine the phase of the reference pulse used to correct the signal. The other heterodyne detector is used to detect the signal. The heterodyne detector efficiency is $\eta_d$ and the detector excess noise $\xi_{d}$. $\gamma$ is the coherent efficiency due to the wavefront aberration of the signal mixing with the LO which is corrected by an AO unit. }
\label{skrlossblock}
\end{figure*}

We present our system model of the D-LLO CV-QKD protocol in Fig.~\ref{skrlossblock}. A Gaussian modulated coherent state (GMCS) of variance $V_A$ is prepared on the satellite (Alice) and measured at the ground station (Bob) using heterodyne detection. At Alice's location in a LEO satellite at altitude $H$, a strong laser source $L_A$ generates pulses (of wavelength $\lambda$, beam-waist $w_0$ and duration $\tau_0$) separated by $2/f$ where $f$ is the repetition rate. A balanced interferometer is used to create self-coherence between signal and reference pulses delayed by $1/f$. The signal is modulated by an amplitude modulator (AM) and phase modulator (PM). The reference and signal pulses are polarization-multiplexed by a polarizing beam-splitter (PBS). After passing through a lossy channel of transmissivity $T$ and channel excess noise $\xi_{ch}$, the reference and signal pulses are de-multiplexed at Bob's side. The signal is further delayed by $1/f$ and a heterodyne detector is used with the LLO. Both signal and reference pulses are received by an aperture of diameter $D_R$. We set the aperture size $D_R$ such that the effects of beam-wandering and elliptical deformation can be neglected. Henceforth, we will assume $T$ is constant and is dominated only by diffraction loss. 

For the LLO, high intensity pulses are generated by the laser $L_B$. These pulses pass through a balanced interferometer to produce self-coherent pulses delayed by $1/f$. The LLO is split by a balanced beamsplitter to the two heterodyne detectors used to measure the quadratures of the reference and signal pulses, respectively. Bob receives the reference pulses which he uses to determine the phase by performing the heterodyne detection using the LLO. Another heterodyne detector is used to measure the quadrature of the signal. The heterodyne detector efficiency of both detectors is $\eta_d$ and the detector excess noise is $\xi_d$. 
The signal wavefront undergoes aberrations by atmospheric turbulence, causing a mismatch at the coherent detection. This is characterized by the coherent efficiency given by%In comparison to the TLO, only the signal and attenuated LO have wavefront aberrations. 
\begin{align}
  \gamma = \frac{|\frac{1}{2} \iint_{\mathcal{D}_R}[ E_\text{LO}^* E_\text{S} +  E_\text{LO} E_\text{S}^*  ] ds|^2}{\iint_{\mathcal{D}_R} |E_\text{LO}| ^2 ds \iint_{\mathcal{D}_R}|E_\text{S}|^2 ds},
  \label{coherente}
\end{align}
where $\mathcal{D}_R$ is the receiver aperture surface, $E_\text{S}$ is the electric field of the signal pulse, and $E_\text{LO}$ is the electric field of the LO that remains undisturbed by the turbulence. 

The coherent efficiency $\gamma$ is effectively the normalized intensity of the wavefront aberration of the signal interfering with the LLO. An AO unit can be inserted to correct the wavefront aberrations of the signal by means of a deformable mirror that is assumed to be controlled faster than the frequency of fluctuations. 

%which is divided by an asymmetrical beamsplitter into the reference and the signal path. 

\begin{table*}[t!]
\centering
\begin{tabular}{| l | l | l | l | l | l | l | l | l | l |}
\hline
$f$& $w_0$ & $D_R$ &$H$ & $\eta_d$ & $V_A$ & $\tau_0$ & $\lambda$ & $\xi_{ch}$\\
\hline
$100$~MHz & $0.15$~m &$1$~m&$500$~km & $0.95$ & $1.5$ & $130$~ps & $1550$~nm &$0.0172$ \\
\hline
\end{tabular}
\caption{System parameters.}
\label{system}
\end{table*}

\subsection{Channel excess noise}
The excess noise is given by
\begin{equation}
\xi=\xi_{ch}+\frac{2 \xi_d}{\eta_d T},
\end{equation} where $\xi_{ch}$ is the channel excess noise which comprises of
\begin{equation}
\begin{split}
\xi_{ch}&=\xi_{ta}+\xi_{RIN, Atmos} +\xi_{Background}\\
&+\xi_{mod}+\xi_{RIN, LO}+\xi_{RIN, Signal}+\xi_{Leak}+\xi_{Phase},
\end{split}
\end{equation} where the terms on the RHS are the time-of-arrival fluctuations $\xi_{ta}$, relative intensity noise (RIN) of RP due to the atmosphere  $\xi_{RIN, Atmos}$, background noise $\xi_{Background}$, modulation noise $\xi_{mod}$, RIN of the LO $\xi_{RIN, LO}$, RIN of the signal due to atmosphere $\xi_{RIN, Signal}$, reference-to-signal leakage $\xi_{Leak}$ and the phase noise after phase correction $\xi_{Phase}$. 
\subsection{Detector excess noise}
The detector excess noise is given by
\begin{equation}
\begin{split}
\xi_{d}&=\frac{2 v_{el}}{\gamma}+\xi_{\gamma}+\xi_{tech},
\end{split}
\end{equation} where the noise contributions listed are the electronic noise $v_{el}$, coherent efficiency noise contribution $\xi_{\gamma}$ and the technical noise $\xi_{tech}$. $\xi_\gamma$ is given by \cite{twang}
\begin{equation}
\xi_{\gamma}=\frac{1-\gamma}{\gamma}.
\end{equation} 
Wavefront aberrations are particularly important in LLO schemes because of the interference of the aberrated signal and un-aberrated LLO wavefront at the detector. In the TLO scheme, both signal and LO are aberrated by the same amount for colinear propagation and this term is $\xi_{\gamma}=0$. 
In the D-LLO scheme, polarization- and time- multiplexing are used to isolate the signal from the reference pulse as in \cite{multiplex}. The remaining photon leakage is given by:
\begin{equation}
\xi_{Leak}=\frac{|\alpha_R|}{R_e+R_{po}},
\end{equation} where $R_e$ is the finite extinction ratio of the pulse generation at Alice and $R_{po}$ is the finite extinction ratio of the polarization beamsplitter (PBS) at Bob. Typical values for $R_e$ and $R_{po}$ are between $30$~dB and $60$~dB.

The technical noise is given by
\begin{equation}
\xi_{tech}=\xi_{ADC}+\xi_{overlap}+\xi_{LO},
\end{equation} where the noise contributions are analogue-digital converter noise $\xi_{ADC}$, detector overlap $\xi_{overlap}$ and LO subtraction noise $\xi_{LO}$. In this D-LLO scheme, a separate heterodyne detector is used to detect the signal. The ADC quantization noise is limited by the maximum amplitude of the signal pulse, instead of the reference pulse as would be the case if one heterodyne detector is used. Subsequently, $\xi_{ADC}=\frac{|\alpha_s|^2}{12 \times 2^n}$, where $|\alpha_s|^2$ is the signal intensity and $n=10$ is the number of bits. Since the signal intensity is at least 2 orders of magnitude smaller than the reference pulse, $\xi_{ADC}$ is negligible. 
\subsection{Phase estimation error} 
In the D-LLO scheme, there are two independent laser sources, one at Alice and one at Bob. The reference pulse $\ket{\alpha_R}$ is sent along with the modulated signal pulse $\ket{\alpha_S}$. %To minimize hardware requirements, the reference and signal pulses are both modulated by the same modulator. 

\begin{table}[b]
\centering
 \begin{tabular}{|l || l|| l|}
 \hline
 Noise term  &  Description & D-LLO \\ [0.2ex]
 \hline\hline
 %$P(T)$ & Channel transmissivity & $20$ $dB$ & \cite{liao} Z \\
     $\xi_{ta}$ & Time-of-arrival fluctuations &  $ 0.0012 V_A$\\
     \hline
               $\xi_{RIN, Atmos}$ & RIN of RP due to atmosphere & $0.002 V_A$\\   
                 \hline     
               $\xi_{RIN, LO}$ & RIN of LO & $0.00035V_A$ \\        
                 \hline
                   $\xi_{RIN, Signal}$ & RIN of signal due to atmosphere &  $<0.0001$ \\             
                     \hline
             %  & $\xi_{Signal, Atmos}$ & RIN of Signal due to atmosphere & Low & Low \\        
         $\xi_{error}$ & Phase estimation error &$\frac{\xi_{ch}+\frac{2(1+\xi_d) }{\eta_d T}}{|\alpha_R|^2} V_A $ \\
           \hline
                     $\xi_{Leak}$ & Photon leakage to signal &$\frac{|\alpha_R|^2}{R_e+R_{po}}$ \\     
                       \hline
         %         $\xi_{drift}$ & Laser drift noise & $0$ & $0$ \\
                           $\xi_{\gamma}$ & Wavefront aberrations & $\frac{1-\gamma}{\gamma}$  \\
                             \hline
       %  $\xi_{AM}$ & Amplitude modulator finite dynamics & $|\alpha_S|^2 10^{-\text{dyn}_{dB}/10}$ &$|\alpha_S|^2 10^{-\text{dyn}_{dB}/10}$ \\      
%& $\eta_d$ & Detector efficiency & $0.6-0.985$ & \cite{vahl} \\
  $\xi_{el}$ & Electronic noise &  $\frac{2 v_{el}}{\gamma}$ \\
    \hline
 $\xi_{tech}$ & Technical noise & $0.005$ \\
% $\xi^{PIR}_{calib}$ & Calibration attack & $\frac{N_0'}{N_0} \xi-2 \mu'+\frac{(\frac{N_0'}{N_0}-1)}{\eta_d T}$& $0$ & $0$\\
%    $\Delta t_j$ & Clock jitter & $100-200$ $ps$ & \cite{xie, anderson} & \\
   %& & $100$ $fs$ &\cite{newbury} & \\
 \hline
 \end{tabular}
 \caption{Excess noise contributions in the satellite-to-Earth channel using the system parameters in Table \ref{system}.}\vspace*{0.3\baselineskip}
 \label{table}
\end{table}

Bob performs a heterodyne detection to determine the phase $\theta_S$ of the signal relative to his LO using the reference pulse.  Bob uses the reference pulse and LO to measure the phase $\theta_R$, and then applies the correction on the signal. After phase compensation, the phase noise for the GMCS protocol $\xi_{phase}$ can be written as \cite{marie}
\begin{equation}
\xi_{Phase}=2V_A (1-e^{-V_{est}/2}),
\end{equation} where $V_{est}$ is the remaining phase error between reference and signal pulse after phase compensation. The phase accumulated by the signal coherent state is
\begin{equation}
\theta_S=\theta^A_{src}+\theta^{ch}_S+\theta_{mod}-\theta^B_{src},
\end{equation} where $\theta^A_{src}$ is the phase of Alice's source, $\theta^{ch}_S$ is the phase introduced by the channel, $\theta_{mod}$ is the modulation phase and $\theta^B_{src}$ is the phase of the LO pulse at Bob's side. The phase of the reference pulse is
\begin{equation}
\theta_R=\theta^A_{src}+\theta^A_{delay}-(\theta^B_{src}+\theta^B_{delay}),
\end{equation} where $\theta^A_{delay}$ and $\theta^B_{delay}$ are the phase delays $1/f$ of the reference pulse and LO, respectively. Note, both the reference and signal are generated at the same source with the phase $\theta^A_{src}$. The remaining phase error $V_{est}=\text{Var}(\hat{\theta}_S-\theta_{mod})$ comprises of
\begin{equation}
V_{est}=V_{error}+V_{drift}+V_{channel},
\end{equation} where $V_{error}$ is the fundamental phase estimation error given by the standard quantum limit:
\begin{equation}
V_{error}=\frac{\xi_{ch}+2\frac{1+\xi_d}{\eta_d T}}{|\alpha_R|^2},
\end{equation} where $\alpha_R$ is the amplitude of the reference pulse prepared by Alice.
Unlike the S-LLO scheme, in the D-LLO scheme, there are two balanced interferometers assuring self-coherence between signal and reference pulses. Consequently, the phase drift noise is eliminated $V_{drift}=0$. Since the reference pulse does not pass through the modulator, the AM dynamics noise component only depends on the signal intensity and therefore, can be neglected.
%The $V_{drift}$ is the phase drift noise given by:
%\begin{equation}
%V_{drift}=2 \pi (\Delta \nu_A+\Delta \nu_B) \frac{1}{f},
%\end{equation} where $\Delta \nu_A$ and $\Delta \nu_B$ are the spectral linewidths of Alice and Bob's laser sources respectively. 
Lastly, the noise of the channel $V_{channel}$ is due to the differences in path length or equivalently the time-of-arrival fluctuations $V_{ta}$ between the signal and reference pulse. 
 %We ignore intrinsic relative intensity noise of the LO as this was demonstrated to be negligible. 

%In the S-LLO scheme, the amplitude modular dynamics must be taken into account. The excess noise due to this effect is given by:
%\begin{equation}
%\xi_{AM}=E^2_{max} 10^{-d_{dB}/10},
%\end{equation} where $E_{max}$ is the maximum amplitude to be modulated to produce the reference pulse. $d_{dB}$ is the ratio between the maximum and minimum amplitudes.
%\subsection{RIN of reference pulse}
%The variation of the reference amplitude will now be derived. The amplitude of the reference pulse is limited by the quantum estimation error $V_{error}$. The estimator can be calculated from the heterodyne measurement outcomes:
%\begin{equation}
%\theta_R=\tan^{-1}\bigg( \frac{p^{(R)}_B}{x^{(R)}_B} \bigg).
%\end{equation}
%The error is
%\begin{equation}
%V_{error}=\text{Var}(\hat{\theta}_R-\theta_R)=\frac{V_B}{V_A (\frac{d (x^{(R)}_B+p^{(R)}_B)}{ d \theta_R})^2}
%\end{equation}
%
%\begin{equation}
%\begin{split}
%V_{error}&=\frac{\chi+1}{2 |\braket{\alpha_R}|^2(1-\frac{ \Delta(\braket{\alpha_R}|)}{|\braket{\alpha_R}|})^2}+\frac{\chi+1}{2 |\braket{\alpha_R}|^2(1+\frac{ \Delta(\braket{\alpha_R}|)}{|\braket{\alpha_R}|})^2} \\
%&\approx \frac{(\chi+1)(1+\sigma^2_{SI})}{|\alpha_R|^2}
%\end{split}
%\end{equation}
%%However, there is an uncertainty of $\sqrt{\text{Var}(\alpha_R)}$ in the amplitude. 
%\subsection{D-LLO}

%We compare the excess noise components of the three system models in the satellite-to-Earth channel, the TLO, S-LLO and D-LLO schemes in Table \ref{table}

\subsection{Optimal reference pulse intensity}
\label{optimal}
In this section, we determine the optimal reference pulse intensity. The most significant difference between noise components in the TLO and LLO is noise $\xi_{LE}$ due to photon leakage from the reference to signal pulses. The larger the reference pulse intensity, the larger $\xi_{LE}$.  However, there is a trade-off with the fundamental quantum phase noise\footnote{Note from this point forward, we make the approximation $\xi_{phase}=2V_A (1-e^{-V_{est}/2})\approx V_A (V_{drift}+V_{error}+V_{ta}+V_{RIN, Atmos})$ and assume that $V_{est}<0.1$.} $\xi_{error}=V_A V_{error}$ which decreases with increasing reference pulse intensity. The reference pulse intensity can be optimized to minimize the excess noise.

%\subsection{S-LLO: AM Dynamics}

%That is,
%\begin{equation}
%\xi=|\alpha_R|^2 10^{-\text{dyn}_{dB}/10}+V_A \frac{\xi_{ch}+2\frac{\eta_d}{T}}{|\alpha_R|^2}+\xi_{other}
%\end{equation} where $\xi_{other}$ are noise contributions that are independent of $|\alpha_R|$. We take the derivative w.r.t. $|\alpha_R|^2$,
%\begin{equation}
%\frac{d\xi}{d |\alpha_R|^2}=10^{-\text{dyn}_{dB}/10}-V_A (\xi_{Ch}+2\frac{\eta_d}{T})/ |\alpha_R|^4=0,
%\end{equation} and rearrange to obtain,
%\begin{equation}
%N_R=|\alpha_R|^2=10^{\text{dyn}_{dB}/20} \sqrt{(\xi_{ch}+2\frac{\eta_d}{T}) V_A}.
%\end{equation} thus
%\begin{equation}
%\xi=2 \times 10^{-\text{dyn}_{dB}/20} \sqrt{V_A (\xi_{ch}+2\frac{\eta_d}{T})}+\xi_{other}
%\end{equation}
%
%For $\text{dyn}_{dB}=40$~dB,  $N_R=100 \sqrt{(\chi_{ch}+1)V_A}$ and thus $\xi=0.02 \sqrt{V_A (\chi_{ch}+1)}+\xi_{other}$. For the same $\text{dyn}_{dB}$, compared to \cite{marie}, $N_R=20 V_A$ and $\xi=0.002 V_A+(\chi_{ch}+1)/20+\xi_{other} \ge 0.05$. 
%
%In the system we consider, the linewidths of the two lasers are $100$~Hz and $10$~KHz respectively at $100$~MHz repetition rate. Thus the phase drift noise of the lasers is $\xi_{drift}=V_{drift} V_A=0.0062 V_A$. 

The photon leakage contributes the noise component $\frac{|\alpha_R|^2}{R_e+R_{po}}$, such that
\begin{equation}
\xi_{ch}=\frac{|\alpha_R|^2}{R_e+R_{po}}+V_A \frac{\xi_{ch}+2\frac{1+\xi_d}{\eta_d T}}{|\alpha_R|^2}+\xi_{other},
\end{equation} with the derivative w.r.t. $N_R=|\alpha_R|^2$ given by,
\begin{equation}
\frac{d\xi_{ch}}{d N_R}=\frac{1}{R_e+R_{po}}-V_A (\xi_{ch}+2\frac{1+\xi_d}{\eta_d T})/ N_R^2=0,
\end{equation} from which it follows that the optimal value for the reference pulse intensity as prepared by Alice (i.e. $T=1$) is
\begin{equation}
N_R=\sqrt{(R_e+R_{po})(\xi_{ch}+2\frac{1+\xi_d}{\eta_d}) V_A},
\end{equation} and the minimum excess noise,
\begin{equation}
\xi_{ch}=2 \times \sqrt{\frac{V_A}{R_e+R_{po}} (\xi_{ch}+2\frac{1+\xi_d}{\eta_d})}+\xi_{other}.
\label{optimal1}
\end{equation}
For detector efficiency $\eta_d=0.95$, $V_A=1.5$, $R_e=60$~dB and $R_{po}=30$~dB (see reference \cite{multiplex}) the initial reference pulse intensity generated by Alice is $N_R\approx 55,000$. For the rest of the paper, we use these parameters to minimize $\xi_{ch}$. We summarize the excess noise contributions in Table \ref{table}.

\section{Numerical simulation of the coherent efficiency}
\label{experiment}

Using the system parameters shown in Table \ref{system} we use numerical methods to obtain the field of the signal involved in calculating the values of $\gamma$ (equation (\ref{coherente})). The numerical methods consist of the use of Fourier algorithms to simulate laser-beam propagation through a turbulent atmosphere. The evolution of the laser-beam is simulated using the open-source software {\it PROPER} \cite{proper}, and the turbulent atmosphere is modelled using phase screens in combination with several atmospheric models.  

To account for the use of an AO system, we assume the existence of hardware, i.e. a deformable mirror, which can apply a correction to each pulse. Each AO correction is represented in the basis of the Zernike polynomials, where the effectiveness of AO ultimately depends on the maximum order, $n_{max}$, of polynomials used to construct each correction. Higher values of $n_{max}$ yield higher values of $\gamma$. A detailed description of the numerical methods used can be found in \cite{eduardo}.  In Table \ref{gamma} we show the resulting mean values of $\gamma$ ($10,000$ iterations were used), with and without AO, for $\zeta=0^o$, and for $\zeta=60^o$. When AO is used, an order $n_{max}=14$ is considered.

%For the system parameters in Table \ref{system} corresponding to a LEO satellite at the zenith angle $\zeta=0$ and receiver aperture of $D_R=1$~m, the wavefront aberrations can be simulated using the PROPER software. An explanation of how this can be done is in \cite{eduardo}. The effectiveness of the AO system is characterized by the maximum order $n_{max}$ of the Zernike polynomials. Higher values of $n_{max}$ correspond to better values of $\gamma$. For the rest of the paper, we set $n_{max}=14$. In Table \ref{gamma}, the results of the simulation for the mean values of $\gamma$ at $\zeta=0^o$ and $\zeta=60^o$, with and without AO are shown ($10,000$ iterations were used). 

\begin{table}[b!]
\centering
\vspace*{1.0\baselineskip}
\begin{tabular}{| l | l | l |}
\hline
$\zeta$ & $\gamma$ &  $\gamma$ with AO \\
\hline
$0^o$ & $0.484$ & $0.843$ \\
\hline
$60^o$ & $0.375$ & $0.677$\\
\hline
\end{tabular}
\caption{Coherent efficiency $\gamma$ in the satellite-to-Earth channel.}\vspace*{1.5\baselineskip}
\label{gamma}
\end{table}

\begin{figure}[b!]
\centering
\includegraphics[scale=0.476]{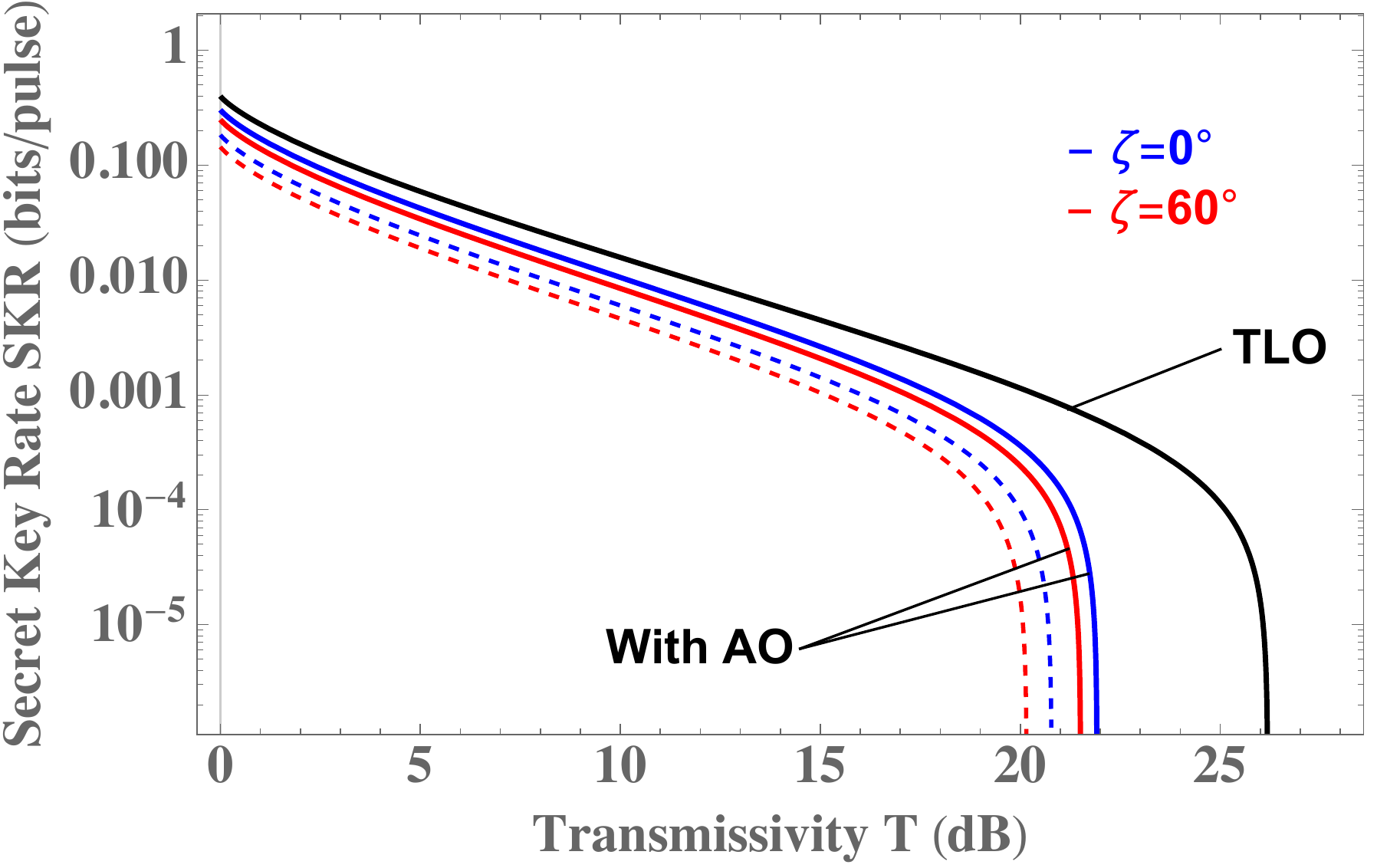}
\caption{Secret key rate under general attacks in the finite-size regime for $V_A=1.5$ comparison for the coherent efficiencies in the satellite-to-Earth channel from Table \ref{gamma}. %The shaded orange region is the secure region at which Eve can't perform a full intercept-resend attack for a realistic value of $N_0'/N_0=1.5$. The region is smaller if $\mu=0.9$ and Bob is unaware of the attack. We compare the two LLO schemes, the S-LLO (blue) and the D-LLO (orange). For the LLO schemes,  
The optimal value for the reference pulse intensity to minimize excess noise is used.}
\label{calibration}
\end{figure}

\section{Practical LLO secret key rate in the satellite-to-Earth channel}
For the Gaussian modulated coherent state protocol with heterodyne detection, the secret key rate\footnote{When we refer to ``secret key rate" in this article, we actually mean a lower bound on the rate.} under general attacks in the deployable setting of the finite regime
%\footnote{For more details, \cite{neda2020} outlines a procedure to calculate $K$ in the finite limit, based on the works of \cite{levc, lupo}.}
(in bits/pulse) is given by \cite{lupo}
\begin{equation}
K= \frac{n}{N} [\beta I_{AB}-S^{\epsilon_{PE}}_{BE}]-\frac{\sqrt{n}}{N}\Delta_{AEP} (n)-\frac{2}{N} \log_2{\frac{1}{2 \epsilon}},
\label{finitekey}
\end{equation} where $I_{AB}$ is the mutual information between Alice and Bob, $0 \le \beta \le 1$ is the reconciliation efficiency, $S^{\epsilon_{PE}}_{BE}$ is the upper bound of the Holevo information taking into consideration the finite precision of the parameter estimation, $N$ is the total number of symbols sent, and $n=N-n_e$, where $n_e$ is the number of symbols used for parameter estimation. $\Delta_{AEP}(n)$ is given by \cite{levc, lupo}
\begin{equation}
\begin{split}
\Delta_{AEP} (n)&=(d+1)^2+4(d+1) \sqrt{\log_2(2/\epsilon_s)}\\
&+2\log_2(2/(\epsilon^2 \epsilon_s))+4 \epsilon_s d/(\epsilon \sqrt{n}),
\end{split}
\label{finitelog}
\end{equation} where $d$ is the discretization parameter\footnote{$d$ is the bits of precision encoded by the symbol. In this work, we set $d=5$ as in \cite{lupo}. }, $\epsilon_s$ is a smoothing parameter corresponding to the speed of convergence of the smooth min-entropy, and $\epsilon_{PA}$ is the failure probability of the privacy amplification procedure. The parameters $\epsilon_s$ and $\epsilon_{PA}$ can be optimized computationally \cite{lev}. In the finite-size regime, one is limited to $\epsilon$-security where $\epsilon=\epsilon_{EC}+2\epsilon_s+\epsilon_{PA}+\epsilon_{PE}$ is the total failure probability of the protocol, and where $\epsilon_{EC}$ is the failure probability of the error correction. 
%It was shown in \cite{lev} that the final secret key rate does not depend strongly on the value of these parameters as (\ref{finitelog}) is logarithmic.

Based on the equations for $I_{AB}$ and $S^{\epsilon_{PE}}_{BE}$ in \cite{kish1}, we calculate the secret key rate against general attacks in the finite-size regime with the block size $n=10^{12}$ , $n/N=0.5$, $\beta=0.95$ and failure of probability (for general attacks) $\epsilon=10^{-55}$. We use the trusted model in which the channel excess noise is untrusted and the detector excess noise is trusted. We use the values of the channel excess noise contributions in Table \ref{table} which are obtained for our system model in the satellite-to-Earth channel. We have used the values from \cite{kish1} to determine the variances $V_{ta}$ and $V_{RIN, Atmos}$ and hence the excess noise contributions $\xi_{ta}=V_{ta} V_A$ and $\xi_{RIN, Atmos}=V_{RIN, Atmos} V_A$, respectively. We note that the time-of-arrival fluctuations contribution would be unchanged for the D-LLO with the exception that it is physically the timing between RP and signal. The reference pulse and signal intensity fluctuates due to the atmospheric turbulence, which adds the same amount of excess noise contribution $\xi_{RIN, Signal}$ and $\xi_{RIN, Atmos}$ as would be the case for the TLO scheme. Similarly, the intrinsic RIN of the LO $\xi_{RIN, LO}$ remains the same. Next, we use the value of (\ref{optimal1}) for the given $R_e=60$~dB and $R_{po}=30$~dB. The electronic noise is set to $v_{el}=0.01$ and technical noise $\xi_{tech}=0.005$. 

In Fig. \ref{calibration}, we plot the secret key rate under general attacks in the finite-size regime versus the transmissivity in units of dB (i.e. $-10 \log_{10}{T}$) at the zenith angles $\zeta=0$ and $\zeta=60^o$. The RP intensity is optimized to minimize the excess noise, and similarly $V_A=1.5$ is chosen to maximize the secret key rate. For the zenith angle\footnote{We take the zenith angle $\zeta=60^o$ to be the worst case scenario. In deployment, the satellite likely spends more time over the duration of the communication link at $\zeta<60^o$.} $\zeta=60^o$, we find non-zero key rates of the D-LLO scheme up to a channel loss of $20$~dB without AO and $22$~dB with AO.

We also plot the key rate of the TLO scheme for comparison. For the TLO scheme, we used the noise model in \cite{kish1} and the same system parameters. Evidently, the TLO performs much better overall and is feasible up to channels losses of $26$~dB. However, the D-LLO scheme without AO is still feasible for channel losses up to $20$~dB ($22$~dB with AO) which is readily achievable for diffraction dominated channel losses of $15$~dB with transceiver aperture diameter $D_T=0.3$~m, receiver aperture diameter $D_R=3$~m and far-field divergence of $10$~$\mu rad$ \cite{kish1}.

\label{secretkeyrate}

\section{Conclusion}
\label{conclusion}
In this work, we developed a practical noise model of a current state-of-the-art LLO scheme (the D-LLO scheme) in the satellite-to-Earth channel. We  numerically simulated the coherent efficiency characterizing the wavefront aberration due to atmospheric turbulence in the satellite-to-Earth channel. Next, we
calculated the expected secret key rates under general attacks in the deployable setting of the finite-size regime, showing that non-zero key rates can be obtained in diffraction-dominated satellite-to-Earth channels. In addition, we found that AO can reduce the excess noise to the point that an observable improvement of the key rates is forthcoming. In conclusion, we find that CV-QKD with an LLO in the satellite-to-Earth channel is indeed feasible. This work extends the scope of our previous work on a TLO scheme to an LLO scheme – the latter providing more security against practical attacks.%\footnote{This paper was submitted to 2021 IEEE International Conference on Communications (ICC): SAC Quantum Communications and Computing Track.}

\section*{Acknowledgment}
This research was funded through a Quantum Technologies Research Network Grant through the Defence Science \& Technology Group, Australian Government.
\bibliographystyle{IEEEtran.bst}

%\bibliography{bib}

\begingroup
\raggedright
\bibliography{IEEEabrv,bib2}
%\bibliography{bib2}
\endgroup

\end{document}